\documentclass[fleqn,12pt,twoside]{article}
\usepackage[headings]{espcrc1}

\readRCS
$Id: espcrc1.tex,v 1.2 2004/02/24 11:22:11 spepping Exp $
\usepackage{graphicx}
\usepackage[figuresright]{rotating}

\newcommand{\AmS}{{\protect\the\textfont2
  A\kern-.1667em\lower.5ex\hbox{M}\kern-.125emS}}

\title{Heavy flavor production in p+p and d+Au collisions at $\sqrt{s_{NN}}=200$ GeV\\ from single leptons over a wide kinematic range
}

\author{Y.~Kwon\address{Department of Physics and Astronomy,\\ 
        401 Nielsen Physics Building, 
        Knoxville, TN 37996-1200, USA}%
         \ for the PHENIX Collaboration
        }
       
\runtitle{Heavy flavor production in p+p and d+Au collisions by PHENIX}
\runauthor{Y. Kwon for PHENIX}

\begin{document}

\maketitle

\begin{abstract}
The PHENIX experiment at the Relativistic Heavy Ion Collider 
has measured single electrons at midrapidity
and 
single muons at forward and backward rapidities ($1.2<|y|<2.2$) 
for p+p and d+Au collisions at $\sqrt{s_{NN}}=200$ GeV. 
The inclusive lepton spectra exhibit excesses over light hadron sources. 
The excesses are attributed 
to semileptonic decays of charm and bottom hadrons 
and compared to the perturbative QCD predictions.
The NLO pQCD calculation underpredicts
the excess lepton production in {\it p+p} collisions. 
The excess lepton production in {\it d+Au} collisions scales
with $N_{coll}$ at midrapidity,
and shows suppression/enhancement relative to the same scaling 
in d/Au-going direction within large uncertainties. 
\end{abstract}

\section{Introduction}

The parton model combined with the perturbative Quantum ChromoDynamics (pQCD)
is applied to violent strong interaction processes 
characterized by a large scale. 
Heavy quark production belongs 
to the category of violent strong interaction processes
due to its large mass
serving as the scale~\cite{pub:factor}. 
The parton model  
is often applied to charm quark production
while classification of the charm quark as a heavy quark is 
not yet fully settled.
Predictions for charm hadroproduction exist from lower collision energies 
including the FNAL fixed target program~\cite{pub:charm2}.
Naive application of parton model to ion collisions, 
i.e. interaction of point-like particles making up nucleons, 
results in a scaling with the number of colliding nucleon pairs, $N_{coll}$.
Systematic studies of charm production in {\it p+p} and p+nucleus collisions 
have been proposed 
as a sensitive way to measure 
the parton distribution functions in nucleons 
and 
the nuclear shadowing effects~\cite{pub:pdf}.

Semileptonic decays of the heavy quark produce 
hard leptons due to its large mass 
and 
produce a large hard lepton excess over the light hadron sources.
PHENIX determines excess leptons over a wide kinematic range
to address heavy quark production.

\section{Measurements}

\begin{figure}[htb]
\begin{minipage}[t]{80mm}
\includegraphics[width=7cm]{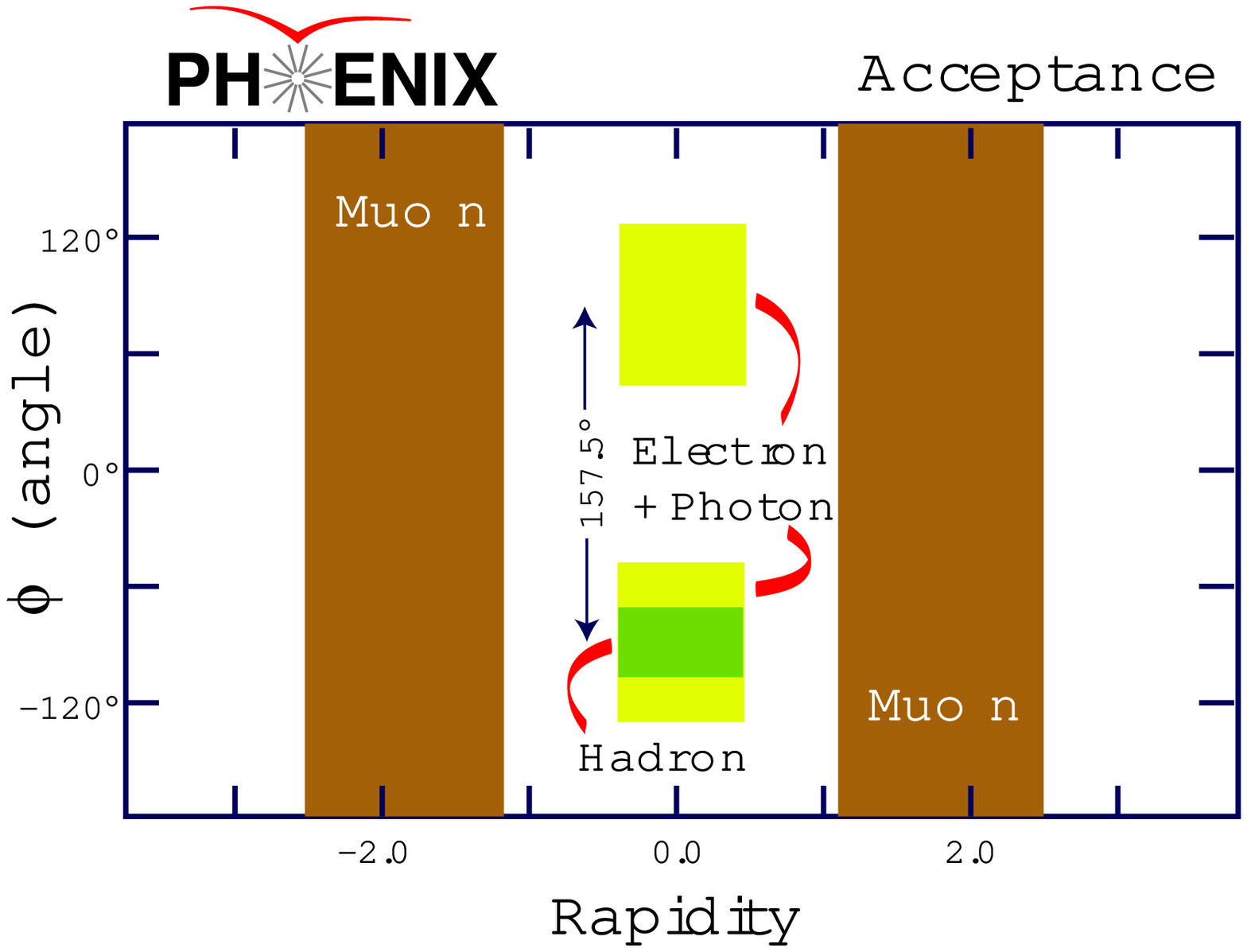}
\caption{Acceptance of the PHENIX experiment}
\label{fig:acceptance}
\end{minipage}
\hspace{\fill}
\begin{minipage}[t]{80mm}
\includegraphics[width=8cm]{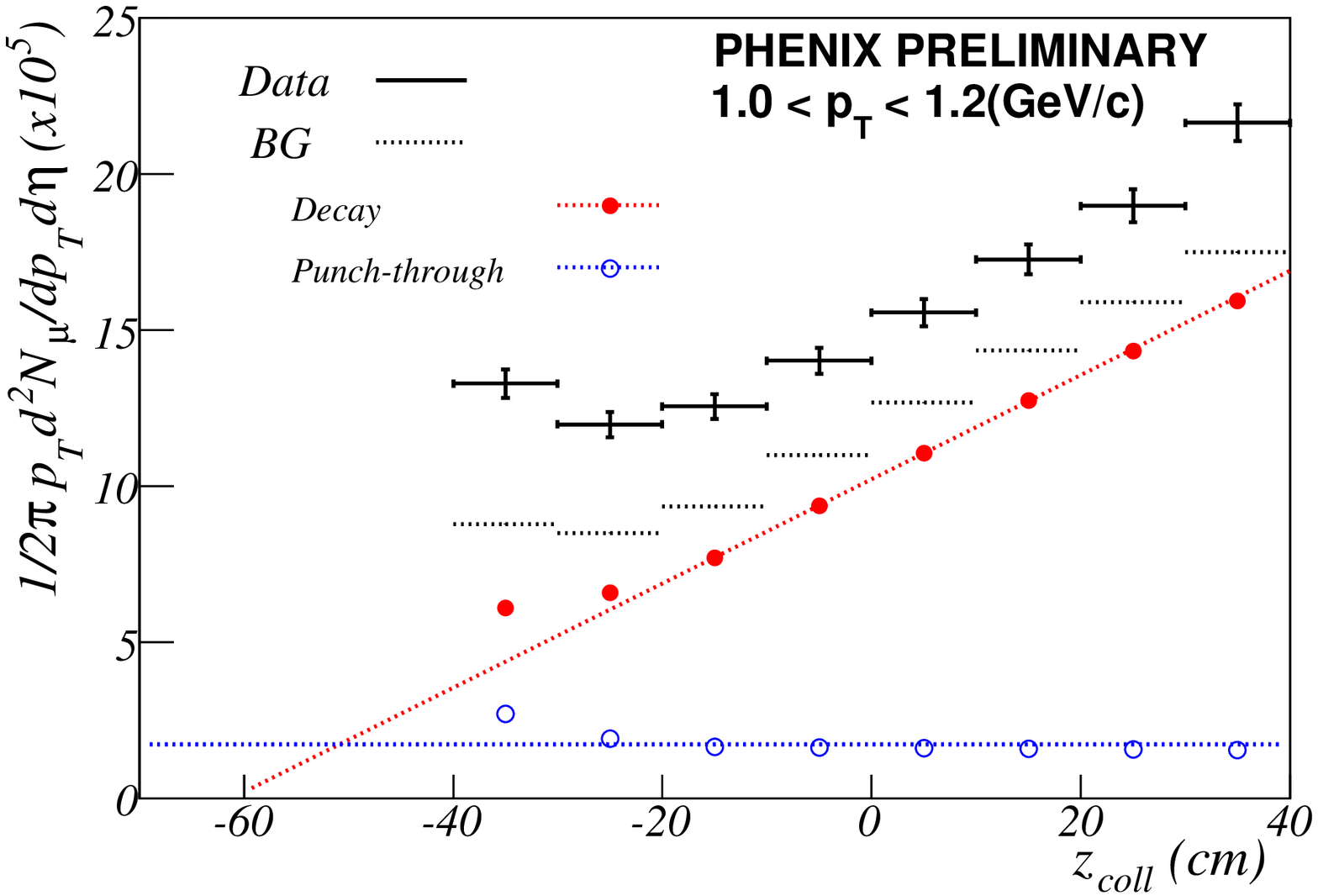}
\caption{Inclusive muon production vs $z_{coll}$.
Light hadron contributions are also shown. }
\label{fig:CompositionOfInclusive}
\end{minipage}
\end{figure}

PHENIX uses global detectors to characterize the collisions,
a pair of central spectrometers at midrapidity to measure electrons,
hadrons, and photons, 
and 
a pair of forward(north)/backward(south) spectrometers to measure muons
(see Fig.~\ref{fig:acceptance})~\cite{pub:detphenix}. 
PHENIX determines the {\it excess leptons}, 
referred to here as 
the {\it non-photonic electrons} 
or
the {\it prompt muons},
by subtracting the contributions by light hadrons 
from the measured inclusive leptons.

\subsection{Non-photonic electrons}  
Inclusive electrons contain two components:
(1) {\it non-photonic} - presumably semileptonic decays of mesons
containing heavy (charm and bottom) quarks, and 
(2) {\it photonic} - Dalitz decays of light neutral mesons 
($\pi^{0}, \eta, \eta^{\prime}, \rho, \omega,$ and $\phi$)
and photon conversions in the detector material. 
PHENIX uses two approaches, 
{\it converter subtraction} and {\it cocktail subtraction},
to estimate the photonic component.  

The converter subtraction is a data-driven approach.
A photon converter (a thin brass tube of 1.7\% radiation length thickness) 
is installed for a special period of runs.
The photon converter multiplies 
the photonic contribution to the electron yield in a well-defined manner
and 
hence the photonic component can be estimated.
The cocktail subtraction simulates electron production 
from the known light hadron sources 
and subtract them from the inclusive spectra.
Dalitz decay of $\pi^{0}$ and photon conversion are 
the dominant source of photonic electron according to the simulation. 
The converter and the cocktail subtraction 
are two independent methods 
and yield consistent results. 

\subsection{Prompt muons}  
Inclusive muons contain three major components:
(1) {\it decay muons} - 
muons from the decays of light hadrons ($\pi$'s and $K$'s),
(2) {\it punch-through's} - 
hadrons that punch through all absorbers and are mis-identified as muons, and
(3) {\it prompt muons} - 
muons produced in the close proximity of collision location 
and 
presumably from the decays of heavy flavor.

Decay muons and punch-through's are separated from the rest 
through their distinct characteristics 
(Fig.~\ref{fig:CompositionOfInclusive}). 
The yield of inclusive muons shows a linear dependence 
on the collision location $z_{coll}$ 
due to muons from the decay of $\pi^{\pm}$'s and $K^{\pm}$'s 
prior to the first absorber material 
located after position along the beam axis $z = -40\ cm$.
We fit the histogram with the function $a + b \cdot z$,
and 
the slope gives the yield per unit length of muons from hadron decay.
The slopes obtained for various transverse momentum $p_{T}$ range are 
described well at low $p_{T}$
and 
constrained at high $p_{T}$
by a data-driven light hadron generator.
Trajectories reconstructed for analysis include partially penetrating ones
as well as completely penetrating muons and punch-through's.
Partially penetrating trajectories are due to the abundant light hadrons,
and 
their yields are linked to the amount of punch-through's 
through the absorber material of known thickness. 
Excess over two sources are identified as due to the prompt muons. 

\section{Result}

The excess lepton production at midrapidity and forward/backward rapidities
is reported for p+p and d+Au collisions at $\sqrt{s_{NN}}=200$ GeV.

\subsection{p+p collisions at $\sqrt{s}=200$ GeV}
\begin{figure}[htb]
\begin{minipage}[t]{82.5mm}
\includegraphics[width=8.25cm]{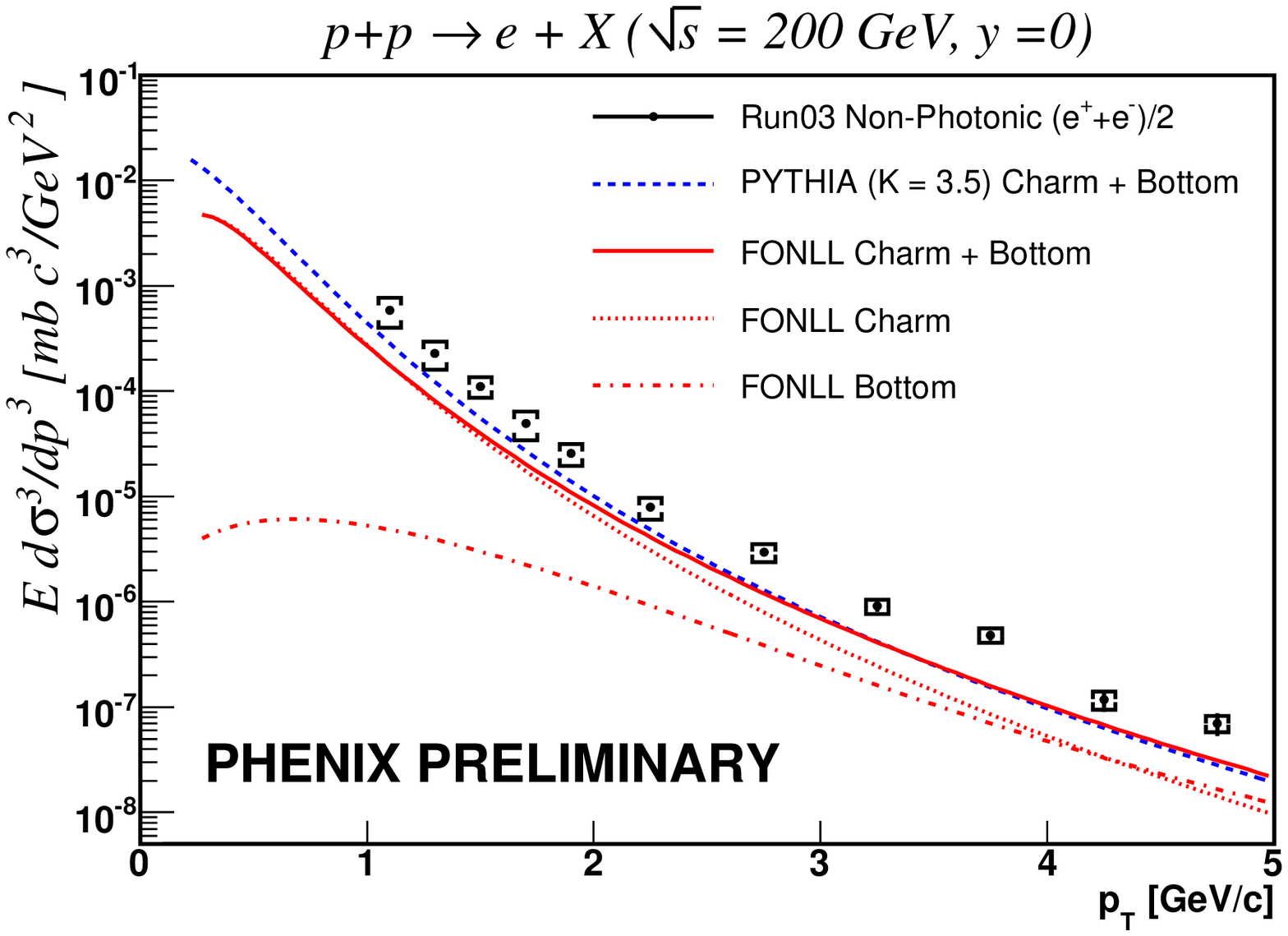}
\end{minipage}
\hspace{\fill}
\begin{minipage}[t]{82.5mm}
\includegraphics[width=8.25cm]{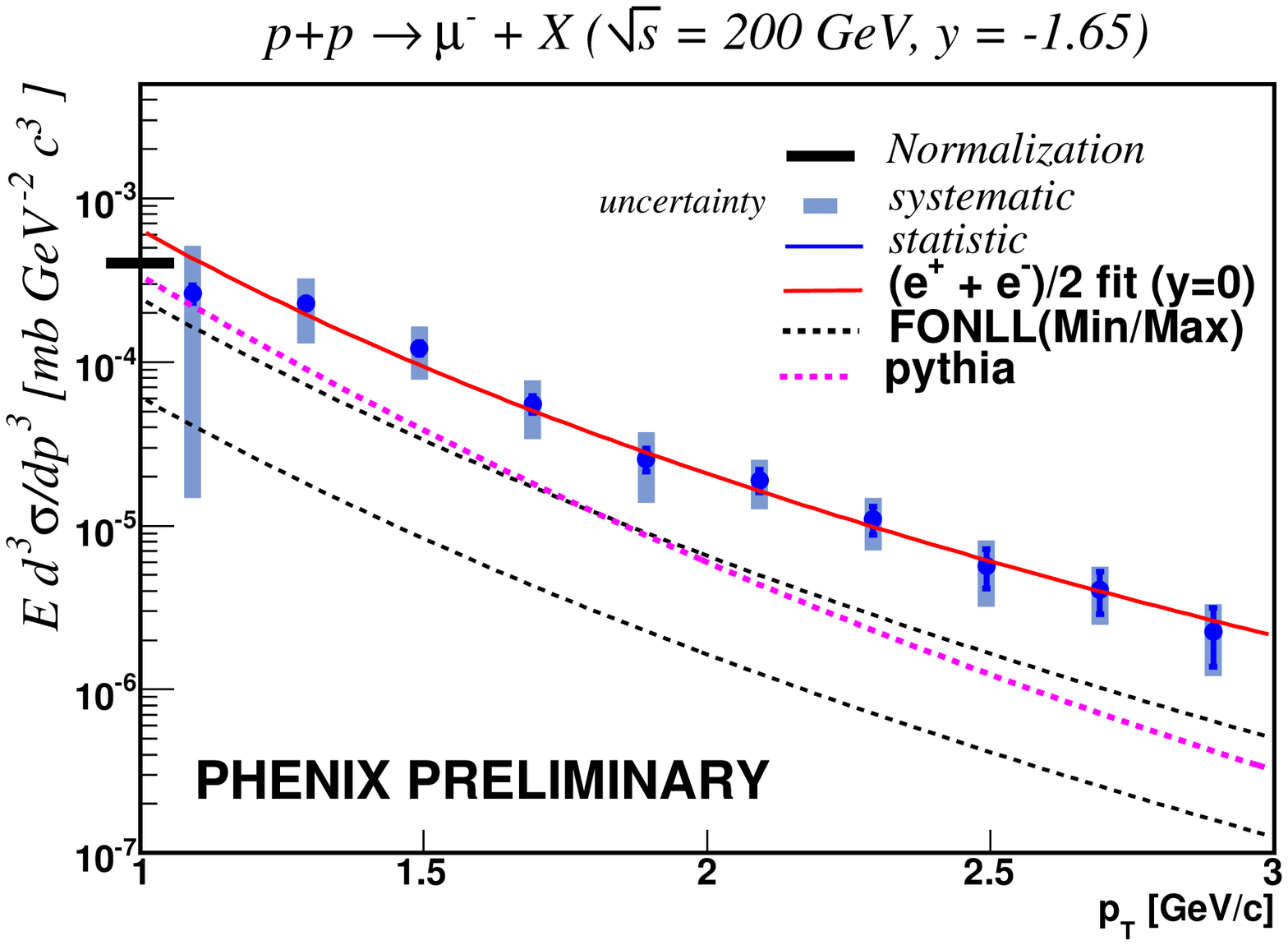}
\end{minipage}
\caption{
Excess lepton production
with PYTHIA LO and FONLL pQCD calculations.
}
\label{fig:excess_lepton}
\end{figure}

The invariant differential cross section of excess leptons are shown
with PYTHIA 
and 
{\it Fixed-Order plus Next-to-Leading-Log}(FONLL~\cite{pub:FONLL}) calculation
in Fig.~\ref{fig:excess_lepton}. 
The PYTHIA/FONLL calculation includes
heavy quark production via 
LO(Leading Order)/NLO(Next to Leading Order) pQCD calculation
and 
its subsequent decay into leptons.
The calculations underpredict the data at high $p_{T}$
within marginal uncertainties.


The plot for prompt muons includes fit of electron spectra, 
and 
shows the excess lepton spectra at y = 0 and at y = 1.65 
are similar over the observed $p_{T}$ range.
The underprediction of the data by the calculation 
at forward rapidity is even stronger than the one at midrapidity
due to the stronger rapidity dependence of production in the calculations.

\subsection{d+Au collisions at $\sqrt{s_{NN}}=200$ GeV}

\begin{figure}[htb]
\begin{minipage}[t]{82.5mm}
\includegraphics[width=8.25cm]{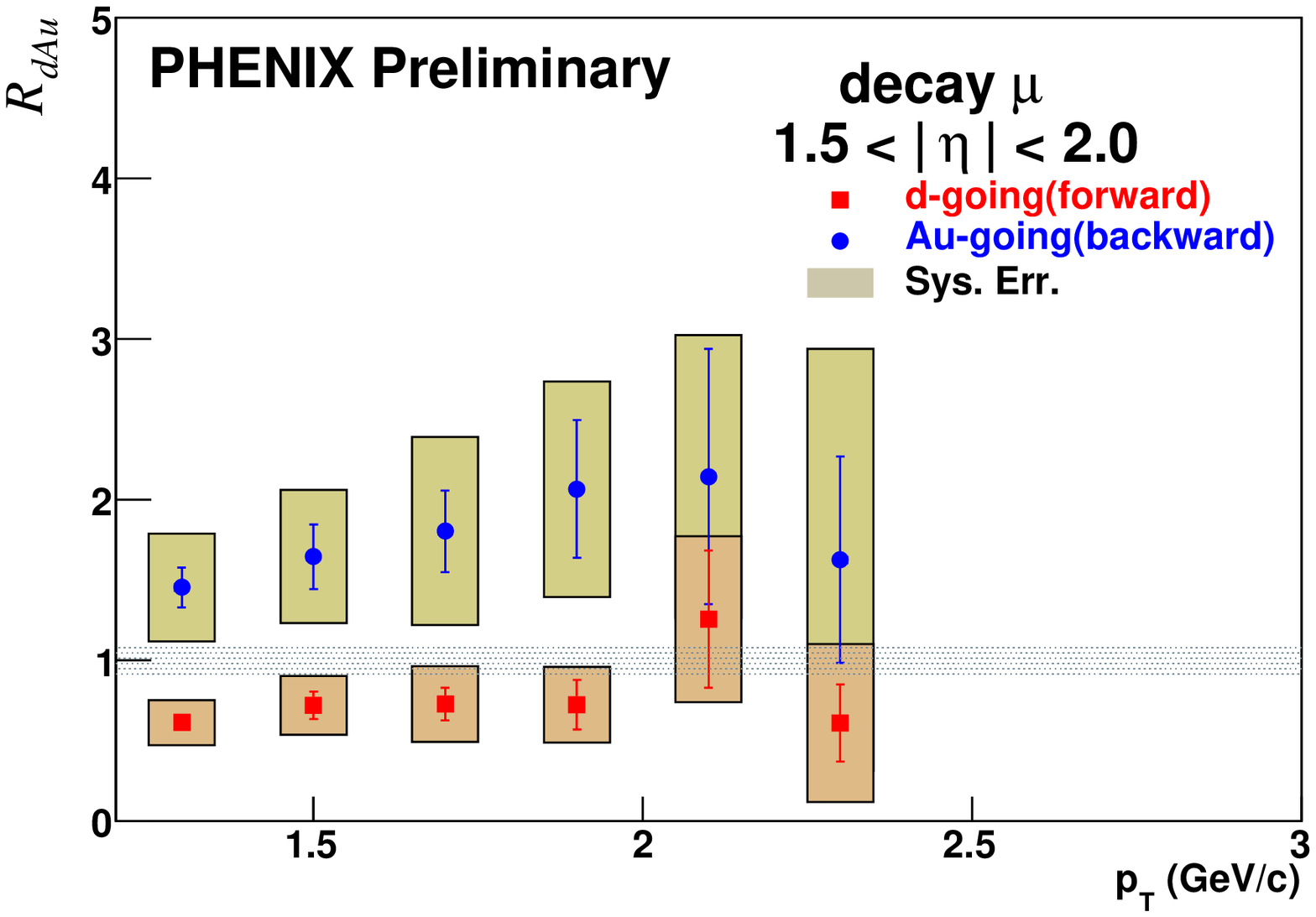}
\end{minipage}
\hspace{\fill}
\begin{minipage}[t]{82.5mm}
\includegraphics[width=8.25cm]{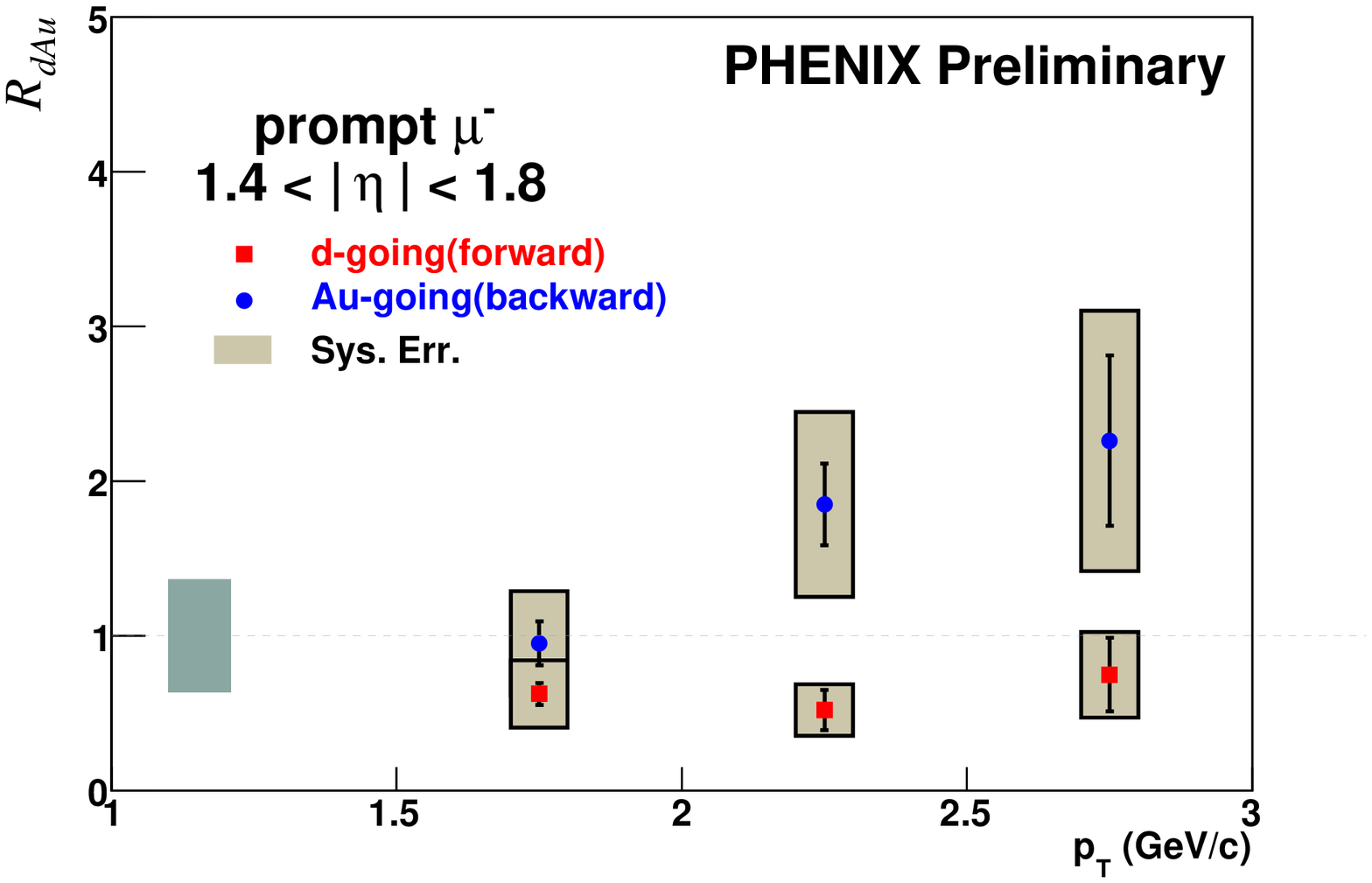}
\end{minipage}
\caption{$R_{dAu}$ for the decay and the prompt muons, d+Au collisions at $\sqrt{s_{NN}}=200$ GeV}
\label{fig:R_dAu_pt}
\end{figure}

The non-photonic electron production in {\it d+Au} collisions 
exhibits scaling with $N_{coll}$ 
indicating point-like interactions~\cite{pub:qm04e}.
Fig.~\ref{fig:R_dAu_pt} shows nuclear modification factor defined as 
\begin{equation}
R_{dAu}(p_{T},\eta) 
\equiv 
( \frac{1}{2 \cdot 197} \cdot
\frac{d^{2}\sigma^{d+Au \rightarrow \mu+X}}{dp_{T}d\eta})/
(\frac{d^{2}\sigma^{p+p \rightarrow \mu+X}}{dp_{T}d\eta})
\end{equation}
for the decay and the prompt muons.
$R_{dAu}$ for d-going(forward)/Au-going(backward) side are 
smaller(suppression)/larger(enhancement) than 1 ($N_{coll}$ scaling) 
within large uncertainties.

\section{Summary}

PHENIX measured excess leptons over light hadron sources
for p+p and d+Au collisions.
FONLL, current NLO pQCD calculation
(semileptonic decays of charm and bottom hadrons), underpredicts 
the measured excess for p+p collisions 
though within marginal uncertainties.
Efforts to reduce the uncertainties are in progress.
If the observed trend remains,
the most likely source of excess hard leptons is,
within the standard model, 
still semileptonic decay of the heavy quark through virtual $W^{\pm}$
since the production through virtual $\gamma$ implies 
abundance of a hard radiation  
or 
large mass lepton pairs not yet observed. 
We presented $R_{dAu}(p_{T})$ of the decay and the prompt muons 
at forward/backward rapidities for d+Au collisions.
Modification of the $N_{coll}$ scaling is seen at forward/backward rapidities
within large uncertainties.


\begin{thebibliography}{9}
\bibitem{pub:factor}
  J. C. Collins, D. E. Soper and G. Sterman, Nucl. Phys. B {\bf 263}, 
  37 (1986).
\bibitem{pub:charm2}
  M. L. Mangano et al., Nucl. Phys. B {\bf 405}, 507 (1993). 
\bibitem{pub:pdf} 
  Z. Lin and M. Gyulassy, Phys. Rev. Lett. {\bf 77}, 1222 (1996).
\bibitem{pub:detphenix}
  K. Adcox et al. (PHENIX), NIM A {\bf 499} 469 (2003).
\bibitem{pub:FONLL}
  M. Cacciari, P. Nason, R. Vogt, http://arxiv.org/abs/hep-ph/0502203.
\bibitem{pub:qm04e}
  S. Kelly for PHENIX, J. Phys G: Nucl. Part. Phys. 30, S1189(2004) 
\end{thebibliography}
\end{document}